# New Green function approach describing the ferromagnetic state in the simple Hubbard model

**Grzegorz Górski[1], Jerzy Mizia, and Krzysztof Kucab***

[1] Faculty of Mathematics and Natural Sciences, University of Rzeszów, ul. Pigonia 1, 35-310 Rzeszów, Poland

* Corresponding author: e-mail: kkucab@univ.rzeszow.pl, Phone: +48 17 851 8578, Fax: +48 17 851 8661

We consider a Hubbard model with occupation dependent hopping integrals. Using the Hartree-Fock (H-F) approximation and the new Green function approach with inter-site kinetic averages included, we analyze the influence of the correlated hopping on ferromagnetic ordering. The influence of correlated hopping on the nonlocal quasiparticle energies and corresponding **k**-dependent spectral weights is included. In addition we obtain the shift of the spin bands which is a major factor in creating spontaneous ferromagnetism. At some parameters of the model the correlated hopping effect is strong enough to achieve saturated ferromagnetism. This state may be obtained at the asymmetric density of states (DOS) and the Fermi energy located in the region of large spectral weight near the band edge. The results are compared with DMFT-Quantum Monte Carlo calculations and with the Hubbard III approximation which includes the correlated hopping effect.

**1 Introduction** The Hubbard model [1,2] is extensively used to analyze strong electron correlations in the narrow energy bands. The classic form of this model describes the behaviour of narrow band electrons in the presence of single site electron-electron repulsion $U = (ii|1/r|ii)$. In the systems with strong on-site Coulomb correlation $U$ there is a split of the spin band into two sub-bands: a lower sub-band centred around the atomic level $T_0$, and the upper sub-band centred around the level $T_0 + U$. The Hamiltonian of basic Hubbard model has the form

$$H = -\sum_{ij\sigma} t_{ij}^{\sigma} c_{i\sigma}^{+} c_{j\sigma} + T_0 \sum_{i\sigma} \hat{n}_{i\sigma} + \frac{U}{2} \sum_{i\sigma} \hat{n}_{i\sigma} \hat{n}_{i-\sigma}, \quad (1)$$

where the operator $c_{i\sigma}^{+}(c_{i\sigma})$ is creating (annihilating) an electron with spin $\sigma = \uparrow, \downarrow$ on the *i*-th lattice site; $\hat{n}_{i\sigma} = c_{i\sigma}^{+} c_{i\sigma}$ is the electron number operator for electrons with spin $\sigma$ on the *i*-th lattice site. In addition to parameter $U$ the model has the nearest-neighbor hopping integral $t_{ij}^{\sigma}$, which in the classic version [1] was occupation independent. In many materials we expect the nearest-neighbor hopping integral to depend on the occupation of the opposite spin electrons on the *i*-th and *j*-th lattice site. The correlated hopping integral between the *i*-th and *j*-th lattice site, $t_{ij}^{\sigma}$, is expressed by the occupation of sites *i* and *j* in the operator form

$$\begin{aligned}t_{ij}^{\sigma} = &\, t_{ij}^{0}(1-\hat{n}_{i-\sigma})(1-\hat{n}_{j-\sigma}) \\ &+ t_{ij}^{1}\left[\hat{n}_{i-\sigma}(1-\hat{n}_{j-\sigma}) + \hat{n}_{j-\sigma}(1-\hat{n}_{i-\sigma})\right] + t_{ij}^{2}\hat{n}_{i-\sigma}\hat{n}_{j-\sigma}\end{aligned}, \quad (2)$$

where $t_{ij}^{0}$ is the hopping amplitude for an electron of spin $\sigma$ when both site *i* and *j* are empty. Parameter $t_{ij}^{1}$ is the hopping amplitude for an electron of spin $\sigma$ when one of the sites *i* or *j* is occupied by an electron with opposite spin. Parameter $t_{ij}^{2}$ is the hopping amplitude for an electron of spin $\sigma$ when both sites *i* and *j* are occupied by electrons with opposite spin.

Hopping integral ($t_{ij}^{\sigma}$) dependence on electron or hole doping was shown by e.g. Hirsch [3] in the simple case of the Hydrogen molecule $H_2$. He proved that the hopping amplitude ($t_{ij}^{2}$) is smaller than the hopping amplitude ($t_{ij}^{0}$). Hopping amplitude at "half filled band" ($t_{ij}^{1}$) has a value in between $t_{ij}^{0}$ and $t_{ij}^{2}$ ($t_{ij}^{0} > t_{ij}^{1} > t_{ij}^{2}$). The same relationship for hopping integrals $t_{ij}^{0}$, $t_{ij}^{1}$, $t_{ij}^{2}$ was also obtained by Feiner,

Jefferson and Raimondi [4] by mapping the three-band extended Hubbard model representing CuO$_2$ planes in High Temperature Superconductors (HTS) to the effective single-band model. A similar result in mapping the three-band model to the single band model was achieved by Simon and Aligia [5].

The correlated hopping effect was used to describe the HTS [5-7], metal-insulator transition [8-12], charge-density wave [13], spin-density wave [13] and ferromagnetism [9,14-18]. An important class of these models was based on the correlated hopping interaction alone $\Delta t_{ij} = (ii|1/r|ij) \equiv t_{ij}^0 - t_{ij}^1$ [14,15,19-26]. One special feature of the hopping interaction is that even for the symmetrical DOS it breaks the symmetry with respect to the half-filling carrier concentration (see [14,15,19]).

As an example, in the mean-field approximation $\Delta t > 0$ it favours ferromagnetism at $0 < n < 1$ stronger than at $1 < n < 2$, [19]. The exact solution in 1D shows the opposite situation [19].

The exact solution of the Hubbard model was obtained only in some specific cases, e.g. for the one dimensional (1D) system [27,28] or infinite dimension system [29]. For two- and three-dimensional systems only the approximate solutions exist. The most established and logically justifiable approximate solution is the Hubbard III approximation [2]. Unfortunately this approximation did not produce the ferromagnetic ground state [30,31]. The approximation Hubbard I [1] brings the ferromagnetic state only at some concentrations and at strongly asymmetric DOS characteristic for the *fcc* lattice. Therefore further research is being carried out on its extension. The original Hubbard-I and Hubbard-III approximations were based on including only the single site correlations. The DOS in those approximations is centred around the atomic level $T_0$, and the level $T_0 + U$. These energies are spin independent. However, Harris and Lange [32] have shown that the spin dependent shift of the two atomic levels is essential for obtaining spontaneous magnetization. Roth [33] has used the inter-site correlations, e.g. : $\langle c_{i\sigma}^+ c_{j\sigma} \rangle$, $\langle \hat{n}_{i\sigma} c_{i-\sigma}^+ c_{j-\sigma} \rangle$, to extend the Hubbard I approximation. She obtained the energy shift enabling the ferromagnetic state. A similar approach was used by Nolting and co-workers in his spectral density approach (SDA) [34-36].

The inter-site correlations were also used in the models extending the Hubbard III approximation [26,37-40], in which the ferromagnetic enhancement was caused predominantly by bandshift between two bands with opposite spins. The weak ferromagnetic state was obtained only at a strongly asymmetric DOS.

In the present paper we analyze the influence of an occupation dependent correlated hopping integral on the ferromagnetic ordering and we include the inter-site correlations. In the analysis we will use the Hartree-Fock (H-F) approximation and the new higher level Green function approach which includes the inter-site kinetic averages. We show how the correlated hopping integral changes the bandwidth, shape of the DOS, and causes spin dependent bandshift. The results show that the correlated hopping integral enables ferromagnetic transition. Transition to ferromagnetism is possible for the asymmetric DOS. At strong disparity of the hopping integrals, the saturate ferromagnetism is obtained at the Fermi energy located in the region of the large spectral weight near the band edge. This result is similar to the result of Kollar and Vollhardt [21] who used the Gutzwiller approximation applied to the hopping interaction alone.

**2 The model** To analyze Hamiltonian (1), with the hopping parameter depending on occupation through Eq. (2), we use equation of motion for the Green functions in the Zubarev notation [41]

$$\varepsilon \langle\langle A; B \rangle\rangle_\varepsilon = \langle [A,B]_+ \rangle + \langle\langle [A,H]_-; B \rangle\rangle_\varepsilon, \quad (3)$$

where $A$ and $B$ are the fermion operators.
Using Hamiltonian (1) in Eq. (3) it produces the following equation for the Green function $\langle\langle c_{i\sigma}; c_{j\sigma}^+ \rangle\rangle_\varepsilon$:

$$\begin{aligned}\varepsilon \langle\langle c_{i\sigma}; c_{j\sigma}^+ \rangle\rangle_\varepsilon = &\delta_{ij} + (T_0 - \mu)\langle\langle c_{i\sigma}; c_{j\sigma}^+ \rangle\rangle_\varepsilon + U\langle\langle \hat{n}_{i-\sigma} c_{i\sigma}; c_{j\sigma}^+ \rangle\rangle_\varepsilon \\ &- \sum_{\alpha\beta l} \xi_\alpha t_{il}^{\alpha\beta} \langle\langle \hat{n}_{l\sigma}^\beta (c_{i-\sigma}^+ c_{l-\sigma} + c_{l-\sigma}^+ c_{i-\sigma}) c_{i\sigma}; c_{j\sigma}^+ \rangle\rangle_\varepsilon \\ &- \sum_{\alpha\beta l} t_{il}^{\alpha\beta} \langle\langle \hat{n}_{i-\sigma}^\alpha \hat{n}_{l-\sigma}^\beta c_{l\sigma}; c_{j\sigma}^+ \rangle\rangle_\varepsilon \end{aligned}$$

(4)

where we used the notation: $\alpha, \beta = \pm$, $\xi_\pm = \pm 1$,

$$\hat{n}_{i\sigma}^+ \equiv \hat{n}_{i\sigma}, \quad \hat{n}_{i\sigma}^- \equiv 1 - \hat{n}_{i\sigma}, \quad (5)$$

and

$$t_{ij}^{\alpha\beta} = \begin{cases} t_{ij}^0 & \alpha = \beta = - \\ t_{ij}^1 & (\alpha = -, \beta = +) \text{ or } (\alpha = +, \beta = -) \\ t_{ij}^2 & \alpha = \beta = + \end{cases} \quad (6)$$

To solve Eq. (4) one has to consider equations of motion for the higher order Green functions which are present on the right hand side of this equation. This leads to the infinite series of equations, therefore one has to truncate it on some level and to introduce the approximation. Initially we will use only the H-F approximation in Eq. (4) to obtain a sort of guiding solution and subsequently we will develop an approximation which includes the inter-site kinetic correlations in the higher order equations for Green functions.

**2.1 Hartree-Fock approximation** Applying to Eq. (4) the H-F approximation we obtain

$$(\varepsilon - T_0 + \mu - Un_{-\sigma} - S_\sigma)\langle\langle c_{i\sigma}; c_{j\sigma}^+ \rangle\rangle_\varepsilon \\ = \delta_{ij} - \sum_l t_{eff,il}^\sigma \langle\langle c_{l\sigma}; c_{j\sigma}^+ \rangle\rangle_\varepsilon, \quad (7)$$

where $S_\sigma$ is the spin dependent band shift

$$S_\sigma = -\frac{1}{N}\sum_{il}\Big[t_{il}^{ex}\langle\hat{n}_{l\sigma}(c^+_{i-\sigma}c_{l-\sigma} + c^+_{l-\sigma}c_{i-\sigma})\rangle \\ -\Delta t_{il}\langle c^+_{i-\sigma}c_{l-\sigma} + c^+_{l-\sigma}c_{i-\sigma}\rangle\Big], \quad (8)$$

and $t^{\sigma}_{eff,il}$ is the effective hopping integral

$$t^{\sigma}_{eff,il} = t^0_{il} + t^{ex}_{il}\left\{\langle\hat{n}_{i-\sigma}\hat{n}_{l-\sigma}\rangle - \langle c^+_{l\sigma}c_{i\sigma}(c^+_{i-\sigma}c_{l-\sigma} + c^+_{l-\sigma}c_{i-\sigma})\rangle\right\} \\ -\Delta t_{il}\langle\hat{n}_{i-\sigma} + \hat{n}_{l-\sigma}\rangle. \quad (9)$$

The hopping interaction $\Delta t_{il}$ appearing in Eqs. (8) and (9) is defined as

$$\Delta t_{il} = t^0_{il} - t^1_{il} \equiv \Delta t, \quad (10)$$

and the exchange-hopping interaction is given by [15]

$$t^{ex}_{il} = \sum_{\alpha\beta}\xi_\alpha\xi_\beta t^{\alpha\beta}_{il} \equiv t^{ex}. \quad (11)$$

Fourier transform of Eq. (7) to the momentum space gives the following result

$$G_{\mathbf{k}\sigma}(\varepsilon) = \frac{1}{\varepsilon - T_0 + \mu - Un_{-\sigma} - S_\sigma - (\varepsilon_\mathbf{k} - T_0)b_\sigma}, \quad (12)$$

where $b_\sigma = \dfrac{t^\sigma_{eff,il}}{t^0_{il}}$ is the parameter of the bandwidth change.

The quasiparticle density of states is given by

$$\rho_\sigma(\varepsilon) = -\frac{1}{\pi}\mathrm{Im}\left[\frac{1}{N}\sum_\mathbf{k}G^\sigma_\mathbf{k}(\varepsilon)\right]. \quad (13)$$

Using Eq. (12) the perturbed density of states $\rho_\sigma(\varepsilon)$ can be written as

$$\rho_\sigma(\varepsilon) = \frac{1}{b_\sigma}\rho_0\left(\frac{\varepsilon - T_0 + \mu - Un_{-\sigma} - S_\sigma}{b_\sigma}\right), \quad (14)$$

where $\rho_0(\varepsilon)$ is the unperturbed DOS for which we assume

$$\rho_0(\varepsilon) = \frac{1+\sqrt{1-a_1^2}}{\pi D}\frac{\sqrt{D^2-\varepsilon^2}}{D+a_1\varepsilon}, \quad (15)$$

with the asymmetry parameter $a_1$ varying continuously from $a_1 = 0$ corresponding to a symmetric semi-elliptic band (or Bethe lattice) to $a_1 \approx 1$ corresponding to a *fcc* lattice [42]. Parameter $D$ is a half bandwidth of the symmetric semi-elliptic band.

For the spin-dependent average occupation number $n_\sigma$ one can write

$$n_\sigma = \int_{-\infty}^{\infty}\rho_\sigma(\varepsilon)f_\sigma(\varepsilon)d\varepsilon, \quad (16)$$

where $f_\sigma(\varepsilon)$ is the Fermi function.

In this H-F approximation the band shift factor $S_\sigma$ and the effective bandwidth factor are given as

$$S_\sigma = \left(-2t^{ex}n_\sigma + 2\Delta t\right)I_{-\sigma}, \quad (17)$$

$$b_\sigma = 1 + \frac{t^{ex}}{t^0_{il}}\left[n^2_{-\sigma} - I_{-\sigma}(I_{-\sigma} + 2I_\sigma)\right] - 2\frac{\Delta t}{t^0_{il}}n_{-\sigma}, \quad (18)$$

with the parameter of kinetic correlation $I_\sigma$ expressed as

$$I_\sigma \equiv \langle c^+_{i\sigma}c_{j\sigma}\rangle \\ = -\frac{1}{N}\sum_\mathbf{k}\frac{(\varepsilon_\mathbf{k}-T_0)}{D}\int_{-\infty}^{\infty}\left[-\frac{1}{\pi}\mathrm{Im}\,G^\sigma_\mathbf{k}(\varepsilon)\right]f_\sigma(\varepsilon)d\varepsilon. \quad (19)$$

The equations (14)-(19) together with expressions for electron occupation and magnetization

$$n = n_\uparrow + n_\downarrow \quad \text{and} \quad m = n_\uparrow - n_\downarrow, \quad (20)$$

form the base for the numerical analysis of the magnetic ground state of the system.

To reduce the number of free parameters we assume additionally that

$$\frac{t^1_{ij}}{t^0_{ij}} = \frac{t^2_{ij}}{t^1_{ij}} = S < 1. \quad (21)$$

Figure 1a shows the phase diagram for the symmetric DOS corresponding to $a_1 = 0$ [in Eq. (15)] and Fig. 1b is for the asymmetric DOS with $a_1 = 0.7$. As has been mentioned before in general, the hopping integrals $t^0_{ij}$, $t^1_{ij}$, $t^2_{ij}$ fulfil the relationship $t^0_{ij} > t^1_{ij} > t^2_{ij}$. We performed numerical calculations for different values of the parameter $S$.

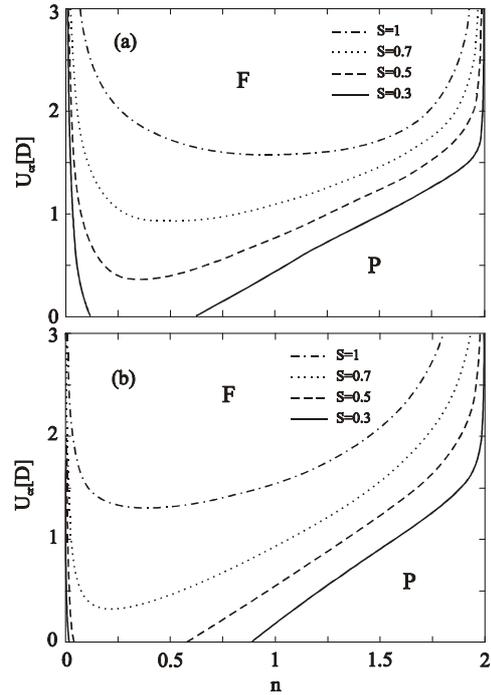

**Figure 1** Dependence of the critical Coulomb interaction on carrier concentration for different values of parameter $S$. Figure 1a is for the symmetric DOS ($a_1 = 0$), and Fig. 1b is for the asymmetric DOS ($a_1 = 0.7$).

A decrease of parameter $S$ (physically, smaller $S$ means a stronger prohibition of hopping between the occupied states) causes a decrease of the critical interaction neces-sary for transition to ferromagnetism, $U_{cr}$. The decrease of $U_{cr}$ is also caused by increasing DOS asymmetry (increasing parameter $a_1$). Figure 2 shows that at small parameter $S$ and for the strongly asymmetrical DOS with a peak at the bottom of the band we obtain ferromagnetism without the Coulomb interaction [15].

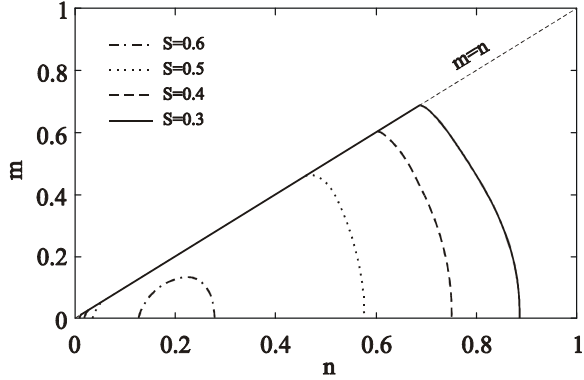

**Figure 2** Magnetization $m$ as a function of carrier concentration $n$ for different values of the parameter $S$. Calculations were carried out in H-F approximation, for the asymmetric DOS ($a_1 = 0.7$) and $U = 0$.

It is apparent that the effect of ferromagnetism was overestimated in this simple H-F approximation, but the results give an insight into the physical effect of gaining the ferromagnetic state with an increasing hopping prohibition parameter in the presence of another electron ($S \searrow$).

**2.2 Green function approach with included inter-site kinetic averages** It is rather obvious from Figure 2 that the H-F approximation enlarges the ferromagnetic ordering beyond its natural limits. Moreover, it is well known that the H-F approximation is not a high quality approximation especially for the systems with strong interaction ($U > D$), where it does not describe the band split and the change of its shape. To explain this effect better we will use the new Green function approach which includes the inter-site kinetic correlations. The equation of motion for the function $\langle\langle \hat{n}_{i-\sigma}^{\alpha} c_{i\sigma}; c_{j\sigma}^{+} \rangle\rangle_{\varepsilon}$ appearing on the right hand side of Eq. (4) can be written in the form

$$\varepsilon \langle\langle \hat{n}_{i-\sigma}^{\alpha} c_{i\sigma}; c_{j\sigma}^{+} \rangle\rangle_{\varepsilon} = \langle \hat{n}_{i-\sigma}^{\alpha} \delta_{ij} \rangle + \varepsilon_{\alpha} \langle\langle \hat{n}_{i-\sigma}^{\alpha} c_{i\sigma}; c_{j\sigma}^{+} \rangle\rangle_{\varepsilon}$$
$$- \sum_{\gamma \beta l} \xi_{\gamma} t_{il}^{\gamma \beta} \langle\langle \hat{n}_{i-\sigma}^{\alpha} \hat{n}_{l\sigma}^{\beta} (c_{i-\sigma}^{+} c_{l-\sigma} + c_{l-\sigma}^{+} c_{i-\sigma}) c_{i\sigma}; c_{j\sigma}^{+} \rangle\rangle_{\varepsilon}$$
$$- \sum_{\gamma \beta l} t_{il}^{\gamma \beta} \langle\langle \hat{n}_{i-\sigma}^{\alpha} \hat{n}_{i-\sigma}^{\gamma} \hat{n}_{l-\sigma}^{\beta} c_{l\sigma}; c_{j\sigma}^{+} \rangle\rangle_{\varepsilon} \quad (22)$$
$$- \xi_{\alpha} \sum_{\gamma \beta l} t_{il}^{\gamma \beta} \langle\langle \hat{n}_{i\sigma}^{\gamma} \hat{n}_{l\sigma}^{\beta} (c_{i-\sigma}^{+} c_{l-\sigma} - c_{l-\sigma}^{+} c_{i-\sigma}) c_{i\sigma}; c_{j\sigma}^{+} \rangle\rangle_{\varepsilon}$$

where

$$\varepsilon_{\alpha} = \begin{cases} T_0 - \mu & \alpha = - \\ U + T_0 - \mu & \alpha = + \end{cases}. \quad (23)$$

The last term in Eq. (22) is neglected since its average is zero. In the fourth term we use the relation

$$\hat{n}_{i-\sigma}^{\alpha} \hat{n}_{i-\sigma}^{\gamma} = \begin{cases} 0 & \alpha = - \\ \hat{n}_{i-\sigma} & \alpha = + \end{cases}, \quad (24)$$

hence

$$\sum_{\gamma} t_{il}^{\gamma \beta} \hat{n}_{i-\sigma}^{\alpha} \hat{n}_{i-\sigma}^{\gamma} = t_{il}^{\alpha \beta} \hat{n}_{i-\sigma}^{\alpha}, \quad (25)$$

what leads to

$$\sum_{\gamma \beta l} t_{il}^{\gamma \beta} \langle\langle \hat{n}_{i-\sigma}^{\alpha} \hat{n}_{i-\sigma}^{\gamma} \hat{n}_{l-\sigma}^{\beta} c_{l\sigma}; c_{j\sigma}^{+} \rangle\rangle_{\varepsilon}$$
$$= \sum_{\beta l} t_{il}^{\alpha \beta} \langle\langle \hat{n}_{i-\sigma}^{\alpha} \hat{n}_{l-\sigma}^{\beta} c_{l\sigma}; c_{j\sigma}^{+} \rangle\rangle_{\varepsilon}. \quad (26)$$

In the above equation we use the following approximation

$$\sum_{\beta l} t_{il}^{\alpha \beta} \langle\langle \hat{n}_{i-\sigma}^{\alpha} \hat{n}_{l-\sigma}^{\beta} c_{l\sigma}; c_{j\sigma}^{+} \rangle\rangle_{\varepsilon} \approx \sum_{\beta l} t_{il}^{\alpha \beta} \langle \hat{n}_{i-\sigma}^{\alpha} \rangle \langle\langle \hat{n}_{l-\sigma}^{\beta} c_{l\sigma}; c_{j\sigma}^{+} \rangle\rangle_{\varepsilon}, \quad (27)$$

and take an average in the third term of Eq. (22) arriving at the following form

$$(\varepsilon - \varepsilon_{\alpha}) \langle\langle \hat{n}_{i-\sigma}^{\alpha} c_{i\sigma}; c_{j\sigma}^{+} \rangle\rangle_{\varepsilon} = \langle n_{i-\sigma}^{\alpha} \delta_{ij} \rangle$$
$$- \sum_{\beta l} t_{il}^{\alpha \beta} \langle \hat{n}_{i-\sigma}^{\alpha} \rangle \langle\langle \hat{n}_{l-\sigma}^{\beta} c_{l\sigma}; c_{j\sigma}^{+} \rangle\rangle_{\varepsilon} \quad (28)$$
$$- \sum_{\gamma \beta l} \xi_{\gamma} t_{il}^{\gamma \beta} \langle \hat{n}_{l\sigma}^{\beta} (c_{i-\sigma}^{+} c_{l-\sigma} + c_{l-\sigma}^{+} c_{i-\sigma}) \rangle \langle\langle \hat{n}_{i-\sigma}^{\alpha} c_{i\sigma}; c_{j\sigma}^{+} \rangle\rangle_{\varepsilon}$$

Equation (28) may be solved by introducing Fourier transforms

$$\langle\langle \hat{n}_{i-\sigma}^{\alpha} c_{i\sigma}; c_{j\sigma}^{+} \rangle\rangle_{\varepsilon} = \frac{1}{N} \sum_{\mathbf{k}} \Gamma_{\mathbf{k},\sigma}^{\alpha}(\varepsilon) \exp\left[ i\mathbf{k} \cdot (\mathbf{R}_i - \mathbf{R}_j) \right], \quad (29)$$

producing the following result

$$(\varepsilon - \varepsilon_{\alpha} - S_{\sigma}) \Gamma_{\mathbf{k},\sigma}^{\alpha}(\varepsilon) = n_{-\sigma}^{\alpha} \left[ 1 + \sum_{\beta} w^{\alpha \beta} (\varepsilon_{\mathbf{k}} - T_0) \Gamma_{\mathbf{k},\sigma}^{\beta}(\varepsilon) \right], \quad (30)$$

where $w^{\alpha \beta} = \dfrac{t_{ij}^{\alpha \beta}}{t_{ij}^{--}}$.

Solution of Eq. (30) has the form

$$G_{\mathbf{k}\sigma}(\varepsilon) = \frac{A_{\mathbf{k}\sigma}^{(1)}}{\varepsilon - \varepsilon_{\mathbf{k}\sigma}^{(1)} - S_{\sigma}} + \frac{A_{\mathbf{k}\sigma}^{(2)}}{\varepsilon - \varepsilon_{\mathbf{k}\sigma}^{(2)} - S_{\sigma}}, \quad (31)$$

where the quasiparticle energies $\varepsilon_{\mathbf{k}\sigma}^{(1)/(2)}$ are given by

$$\varepsilon_{\mathbf{k}\sigma}^{(1)/(2)} = \frac{1}{2} \left[ \varepsilon_{+} + \varepsilon_{-} + (n_{-\sigma}^{-} w^{--} + n_{-\sigma}^{+} w^{++})(\varepsilon_{\mathbf{k}} - T_0) \right]$$
$$\mp \frac{1}{2} \left\{ \left[ \varepsilon_{+} - \varepsilon_{-} + (n_{-\sigma}^{+} w^{++} - n_{-\sigma}^{-} w^{--})(\varepsilon_{\mathbf{k}} - T_0) \right]^2 \quad , \quad (32)$$
$$- 4 n_{-\sigma}^{+} n_{-\sigma}^{-} w^{+-} w^{-+} (\varepsilon_{\mathbf{k}} - T_0)^2 \right\}^{1/2}$$

with their spectral weights

$$A^{(1)}_{\mathbf{k}\sigma} = \frac{\varepsilon^{(1)}_{\mathbf{k}\sigma} - n^-_{-\sigma}\varepsilon_+ - n^+_{-\sigma}\varepsilon_-}{\varepsilon^{(1)}_{\mathbf{k}\sigma} - \varepsilon^{(2)}_{\mathbf{k}\sigma}}$$
$$+ \frac{n^-_{-\sigma}n^+_{-\sigma}\left(w^{+-}+w^{-+}-w^{++}-w^{--}\right)(\varepsilon_{\mathbf{k}}-T_0)}{\varepsilon^{(1)}_{\mathbf{k}\sigma} - \varepsilon^{(2)}_{\mathbf{k}\sigma}}, \quad (33)$$

$$A^{(2)}_{\mathbf{k}\sigma} = 1 - A^{(1)}_{\mathbf{k}\sigma}. \quad (34)$$

The band shift parameter in Eq. (31) has now the form

$$S_\sigma = \frac{t^{ex}}{U}\frac{2}{N}\sum_{\mathbf{k}}\frac{(\varepsilon_{\mathbf{k}}-T_0)}{D}\int_{-\infty}^{\infty}(\varepsilon - \varepsilon_{\mathbf{k}})\left[-\frac{1}{\pi}\mathrm{Im}\,G^\sigma_{\mathbf{k}}(\varepsilon)\right]f_\sigma(\varepsilon)d\varepsilon$$
$$-2\Delta t I_{-\sigma} \quad (35)$$

with $I_{-\sigma}$ given by Eq. (19).

This parameter is the most important factor in the generation of ferromagnetism in the model presented. Its value depends on carrier concentration, magnetization and the hopping correlation factor $S$.

**3 Numerical results of new Green function approach** In the model presented, ferromagnetism is generated by differences in hopping integrals $t^0_{ij}$, $t^1_{ij}$, $t^2_{ij}$ described by parameter $S$ [see Eq. (21)]. Those differences create spin dependent band shift and a change of the spin band width which depends on concentration. These two effects were demonstrated in the H-F approximation and now we will show how they work in the higher order approximation. In the H-F approximation (see Fig. 1) at some model parameters, which evidently overestimate this ordering, they created spontaneous ferromagnetism even without Coulomb repulsion.

Now in Fig. 3 we present the results of the DOS calculations in the paramagnetic state at different values of parameter $S$ in this new higher level Green function approach, which includes the inter-site kinetic averages. The initial (unperturbed) DOS is asymmetrical with $a_1 = 0.7$.

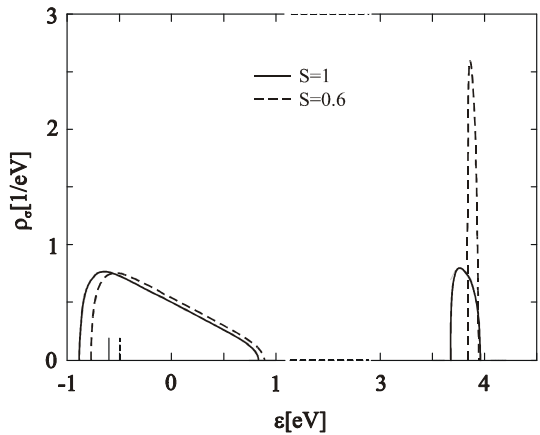

**Figure 3** Electron density of states as a function of energy for different values of the correlated hopping parameter, $S$, in the paramagnetic state ($n = 0.365$, and $m = 0$). In the calculations we used the asymmetrical initial DOS with $a_1 = 0.7$. Other parameters are: $U = 4D$, and $D = 1\mathrm{eV}$.

The lower Hubbard band has very small bandwidth change and large band shift caused by parameter $S_\sigma$ (not shown in this figure). The upper band has a strong bandwidth change under the influence of factor $S$. For $S=1$, which corresponds to $t^0_{ij} = t^1_{ij} = t^2_{ij}$, the maximum DOS in the lower and upper Hubbard band is the same. However, at $S<1$ under the correlation hopping effect the upper band becomes narrow with a high maximum, which is evidence of strong localization.

The band shift parameter $S_\sigma$ has a strong dependency on the correlated hopping parameter $S$ and on electron concentration $n$ (see Fig. 4). In the split band limit, close to the half-filling concentration, the lower band is filling up and the parameter $S_\sigma$ goes to zero. It does not support ferromagnetic ordering therefore this ordering is not possible at this concentration.

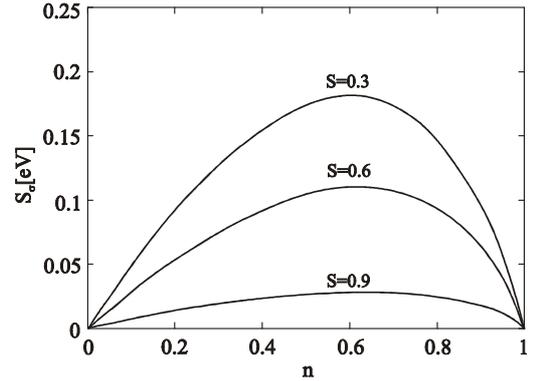

**Figure 4** The band shift factor, $S_\sigma$, in a function of electron concentration for different parameters $S$ in the paramagnetic state; initial DOS with $a_1 = 0.7$, $U = 4D$, and $D = 1\mathrm{eV}$.

In Fig. 5 we show the DOS for $n = 0.365$ and $m = 0.29$ in the presence of small correlated hopping parameter $S = 0.3$ (strong correlation).

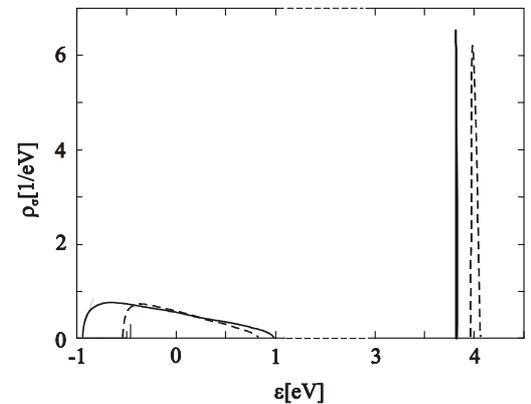

**Figure 5** Electron density of states as a function of energy for majority spin band (solid line) and minority spin band (dashed line) in the ferromagnetic state ($n = 0.365$, and $m = 0.29$). Calculations were done for the strong correlated hopping effect ($S = 0.3$), and asymmetrical DOS with $a_1 = 0.7$. Other parameters are: $U = 4D$, and $D = 1\text{eV}$.

The results show a large shift of centres of gravity between both spin bands. The upper majority spin band does not even overlap with the upper minority spin band. Such a large shift between spin bands enables ferromagnetic ordering at a high kinetic correlation ($S = 0.3$). The differences in the widths of spin bands, especially in the upper Hubbard band are also visible. Shifts and bandwidth changes caused by strongly correlated hopping band are the force behind ferromagnetic ordering.

Figures 6 and 7 shows the dependence of the magnetic moment on carrier concentration for different values of hopping correlation $S$. Both figures are prepared for the asymmetric DOS with different asymmetry parameters.

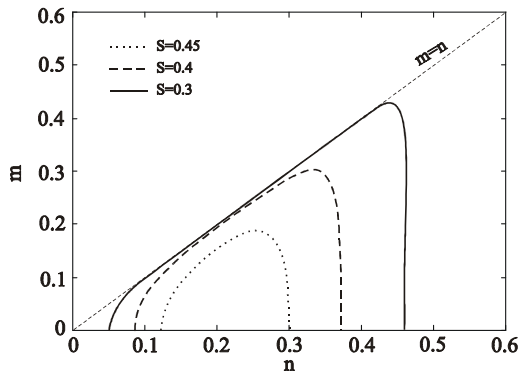

**Figure 6** Magnetization $m$ as a function of carrier concentration $n$ for different values of parameter $S$ and, $U = 4D$. Calculations were performed for the asymmetric DOS ($a_1 = 0.7$), and $D = 1\text{eV}$.

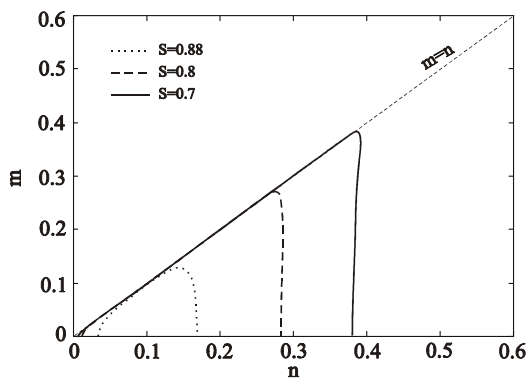

**Figure 7** Magnetization $m$ as a function of carrier concentration $n$ for different values of parameter $S$ and $U = 4D$. Calculations were performed for the strong asymmetric DOS ($a_1 = 0.97$), and $D = 1\text{eV}$.

At strong correlation ($S < 0.4$) we obtain for some values of $n$, saturated ferromagnetism. The range of parameter $S$ for which we have spontaneous ferromagnetism depends on the DOS asymmetry parameter $a_1$. At $a_1 = 0.7$ results in Fig. 6 show that spontaneous ferromagnetism is observed at $S < 0.5$, while at the stronger DOS asymmetry ($a_1 = 0.97$) in Fig. 7 it is observed already at $S < 0.9$.

At $a_1 = 0.7$ and strong hopping correlation ($S < 0.3$) at large carrier concentration (larger than the concentration corresponding to the maximum value of magnetization) we obtain two solutions for magnetization at given $n$. In this region [ $n > n(m_{\max})$ ] magnetization strongly decreases with increasing carrier concentration.

We had studied functional dependence $m(n)$ in the full Hubbard III approximation before [26] where only hopping interaction $\Delta t$ was included (without $t_{ex}$). In this analysis [26], performed with the same DOS and the same asymmetry factor ($a_1 = 0.7$) as in the current paper, the maximum of ferromagnetic ordering was shifted to concentrations larger than $n = 0.5$ with $n(m_{\max})$, growing with increasing interaction $\Delta t$.

In our previous analysis, extending the Hubbard III approximation [26] and CPA approximation [37] the Coulomb on-site correlation had a strong influence on ferromagnetism (see also [16,17]). This correlation in the split band limit and in the case of symmetric DOS favors ferromagnetism near half-filling. Asymmetry of DOS helps only a little in ferromagnetism.

In the present analysis (see Figs 6 and 7) ferromagnetic alignment is located at concentrations $n < 0.5$ since the effect of the Coulomb correlation on ferromagnetism in the Hubbard I approximation is much smaller than in the Hubbard III approximation, where it stimulates ferromagnetism at higher concentrations $0.5 < n < 0.8$ (see [26]).

**4 Conclusions** In this paper we analyze the influence of correlated hopping on the existence of the ferromagnetic state in the H-F approximation and in the Green function decoupling which includes the inter-site kinetic correlations. In the H-F approximation the ferromagnetic state was already obtained at small $U$ and even for the symmetric DOS. This is obviously the result of poor approximation. Nevertheless this approximation has shown two major effects of the inter-site kinetic correlation. These are the bandwidth change described by factor $b_\sigma$ [given by Eq. (18)] and the band shift change given by factor $S_\sigma$ [Eq. (17)]. At a strong enough asymmetry of DOS (when DOS is much alike the *fcc* DOS) and at strong correlated hopping factor $S$ the kinetic effects by themselves generate spontaneous magnetization, even without the potential term i.e. at $U = 0$.

The H-F approximation neglects changes in the band shape and at higher $U$ neglects the splitting of the band into two separated Hubbard bands. Despite its shortcomings this approximation demonstrates a strong influence by the factors $b_\sigma$ and $S_\sigma$ on ferromagnetism.

In the next step we used a higher level, more realistic Green function approximation to see what these two factors look like and how they influence the ferromagnetism. At the symmetric DOS the kinetic correlation supports ferromagnetism only weakly. To obtain this phase one needs the asymmetrical DOS with a relatively high correlation factor (low $S$). There is an interdependence between DOS asymmetry and strength of kinetic correlation described by factor $S$. An increase of parameter $a_1$ (asymmetry) produces ferromagnetic alignment at smaller kinetic correlation. The major factor by which the correlated hopping ($S$) creates ferromagnetism is the band shift. Spin dependent change of the bandwidth has only a small contribution towards ferromagnetic ordering.

As we have seen, ferromagnetism in the Hubbard I like approximation is strongly influenced by the asymmetry of DOS. Such an effect of gaining ferromagnetism by the DOS with very strong asymmetry was reported by Wahle et al. [42] and Ulmke [43], who used quantum Monte Carlo simulations with the same DOS as in Eq. (15) but with stronger a asymmetry factor $a_1 = 0.97$. They had already obtained ferromagnetic ordering at small concentrations where the Fermi level is located near the strong peak in the DOS. Their results resemble ours (see Fig. 7) since the direct Monte Carlo simulation takes care of hopping prohibition in the presence of another electron. This effect in our more analytical approach is implemented by the prohibition factor $S$.

The results obtained in the H-F approximation as well as in the higher level Green function approximation show the importance of the correlated hopping effect in the creation of itinerant ferromagnetism.